\documentclass[12pt]{iopart}
\usepackage{graphicx}

\begin{document}

\title{Condensation transition in a model with attractive particles and non-local hops}

\author{Apoorva Nagar}

\address{Department of Theory and Bio-Systems, Max Planck Institute of Colloids and interfaces}
\ead{apoorva.nagar@mpikg.mpg.de}

\begin{abstract}
We study a one dimensional nonequilibrium lattice model with competing features of particle attraction and non-local hops. The system is similar to a zero range process (ZRP) with attractive particles but the particles can make both local and non-local hops. The length of the non-local hop is dependent on the occupancy of the chosen site and its probability is given by the parameter $p$. Our numerical results show that the system undergoes a phase transition from a condensate phase to a homogeneous density phase as $p$ is increased beyond a critical value $p_c$. A mean-field approximation does not predict a phase transition and describes only the condensate phase. We provide heuristic arguments for understanding the numerical results.   
\end{abstract}

\maketitle

\section{Introduction}

One dimensional lattice models are important tools in nonequilibrium physics. They are used to model physical systems like shaken granular gases~\cite{torok, evans}, vehicular traffic~\cite{nagel} and ribosomal motion on mRNA~\cite{macdonald}. The scope of these models however is larger than modelling particular phenomena. In absence of a general formalism, their study helps us understand the nature of steady states and phase transitions in nonequilibrium systems~\cite{schmittmann}. In particular, the asymmetric simple exclusion process (ASEP)~\cite{ligget} and the zero range process (ZRP)~\cite{spitzer} have emerged as paradigms and have provided us important insights owing to their simplicity and analytic tractability. The ASEP with open boundaries shows the interesting phenomenon of boundary induced phase transitions~\cite{krug}. The phase behaviour and steady state properties in this case can be obtained analytically~\cite{derrida}. Apart from open boundaries, dynamical features also have a significant effect on the steady state. The steady state of the ZRP is known to be a product measure for certain forms of the hop rate, allowing analytic treatment~\cite{evans,evans1}. Other dynamical features like chipping and aggregation~\cite{satya,satya1,rajesh,levine}, disorder~\cite{barma} and evaporation/deposition~\cite{parmeggiani} have been shown to affect the steady state behaviour. Our work involves studying the effect of non-local hops, a feature whose effect on simple exclusion process has recently been explored~\cite{ha, nagar, nossan, vernon, otwinowski}. The study of ZRP like models with hops of arbitrary length has been done in the equilibrium context in the urn models~\cite{godreche}. \\

In this paper we study a one dimensional lattice model similar to the ZRP, allowing for unlimited occupancy. The particles have an attractive interaction with hopping rates inversely proportional to the occupancy $n_i$, $i$ being the label of the site to which they belong. The dynamics consists of two possible moves - a particle from a randomly chosen site can either move to one of the nearest sites with a probability $(1-p)/2$ or make a non-local hop to one of the two sites at a distance of $n_i$ with probability $p/2$ (see Fig.~(\ref{dynamics})). Thus the non-local move is always to a distance equal to the occupancy of the chosen site. While the attractive interaction aids the clustering of particles on a site, the non-local hops promote declustering. We show that the competition between these opposing mechanisms leads to a phase transition between a condensate and a homogeneous density phase. It is well known for the ZRP that if the hop rate  decays to zero in the limit of infinite occupancy, there is a condensate formations at all densities~\cite{evans}. Thus the $p \rightarrow 0$ limit of our model has a steady state with a condensate containing a finite number of particles, while the density on the rest of the lattice approaches zero. As we will see, the steady state changes as $p$ is increased and the condensate disappears at a finite value of $p$. \\           

Our model can be compared to the conserved mass aggregation (CA) models studied in ~\cite{satya,satya1,rajesh} where a phase transition occurs due to the opposing mechanisms of chipping and aggregation. In these models, a cluster of particles moves as a whole and merges with particles already present at the arrival site. This promotes clustering. On the other hand, it is also possible for single particles to chip off from a cluster, promoting declustering. The control parameters here are the density and the probability of chipping. As the density is increased for a given chipping probability, the system goes from a homogeneous to a condensate phase.\\ 

Our results show some similarity with the model described above. We find a phase transition controlled by two parameters - the density and the probability of non-local hops (Phase diagram, Fig.~(\ref{phase})). While nonequilibrium models with a condensation transition have been well studied, the importance of this work lies in a novel physical mechanism for the transition. The melting of the condensate in our case crucially depends on the symmetry in particle dynamics and on the dimension. The phase transition happens only with symmetric dynamics and is expected to occur only in one dimension. We will describe the physical mechanism behind these results in the discussion section.\\

The paper is organised as follows. In section $2$, we describe the model and present numerical results for the phase transition. We analyse these results in section $3$. We find that a mean field approximation for the model does not capture the phase transition. We provide heuristic arguments for the physical mechanism  of the transition. Our arguments are supported by numerics. We summarise our results in section $4$.\\

\section{Description of the model, Numerical results}         

\subsection{The model}

\begin{figure}
\begin{center}
\includegraphics[width=0.8 \linewidth,angle=0]{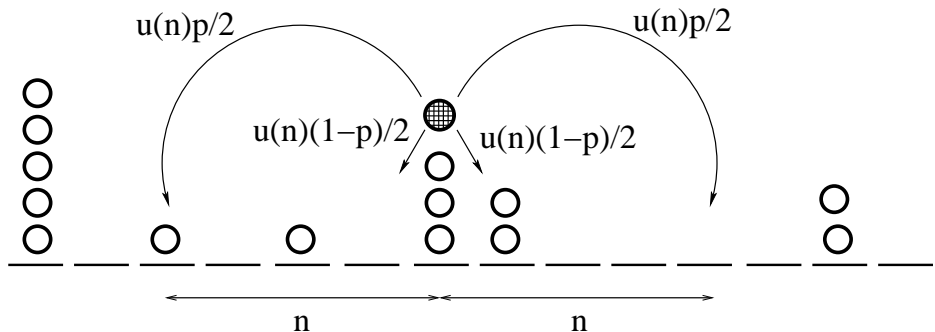}
\caption{Schematic diagram showing the possible dynamical moves. A particle (shown in grey) from a randomly chosen site can attempt one of the four possible moves shown with the corresponding probabilities. The value of $n$ in the above example is $4$.}
\label{dynamics}
\end{center}
\end{figure}

We consider a lattice with $L$ sites and $N$ particles moving on it. We take periodic boundary conditions and allow for arbitrary occupancy at the sites. The dynamics is as follows: we choose a site at random and if it is occupied, we attempt to move a particle from this site. The rate of a particle moving out of this site, $u(n)$, is determined by the occupancy $n$ of the site and we take $u(n)=1/n$. A particle that leaves the site can either make a local move to one of the nearest neighbour sites with probability $(1-p)/2$, or it can make a non-local move to one of the two sites at a distance $n$ with probability $p/2$. Here $p$ lies in the interval $[0,1]$. The dynamical moves with the overall rates of particle motion are shown schematically in Fig.~\ref{dynamics}. The non-local hops and periodic boundary condition impose a limit on the number of particles. Long jumps of length $L/2$ or more will move the particle closer to the originating site instead of taking it farther. Therefore, the maximum possible value of $N$ is $\frac{L}{2}-1$.\\

\subsection{Numerical results}

\begin{figure}
\begin{center}
\includegraphics[width=0.33 \linewidth,angle=-90]{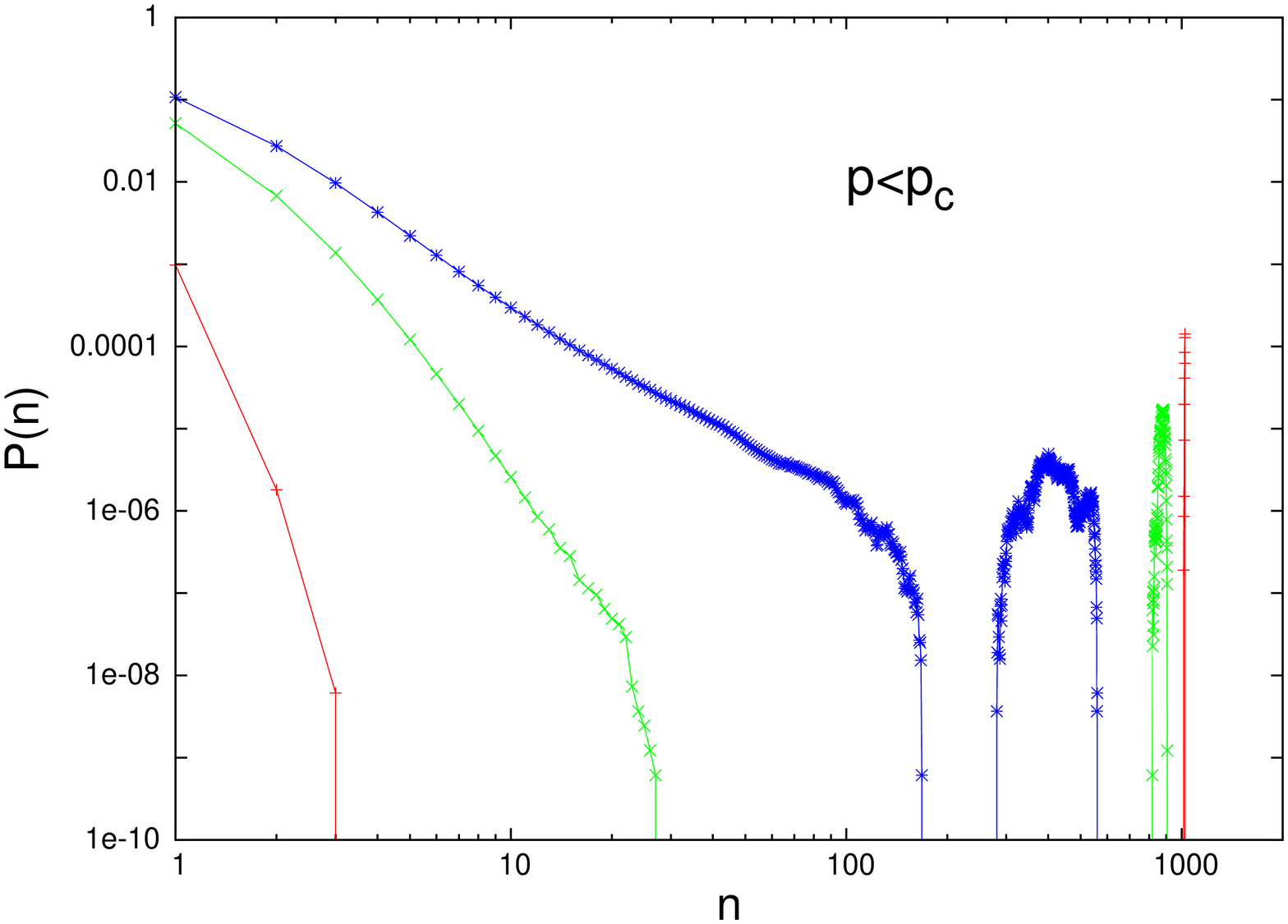}
\includegraphics[width=0.33 \linewidth,angle=-90]{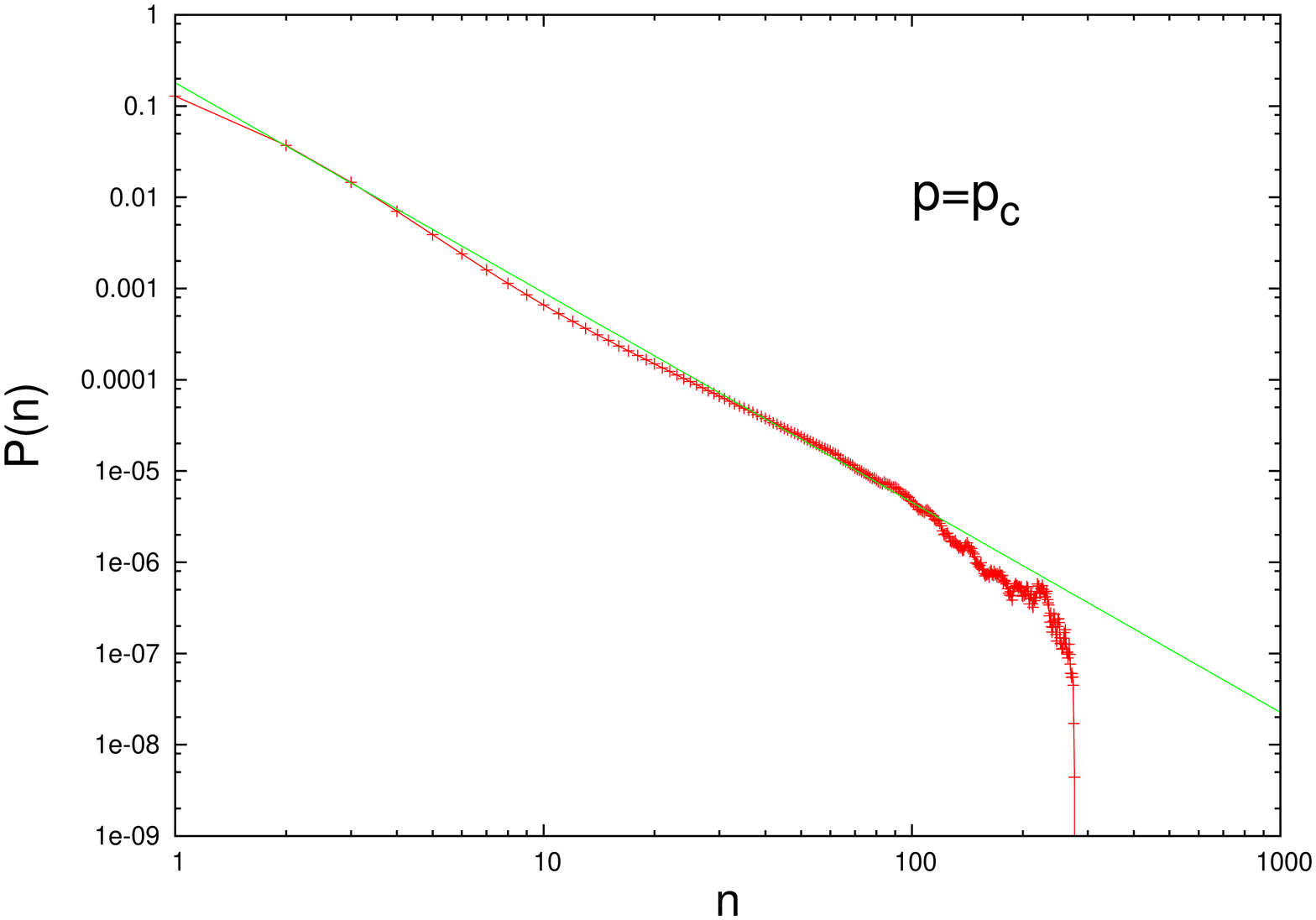}
\includegraphics[width=0.33 \linewidth,angle=-90]{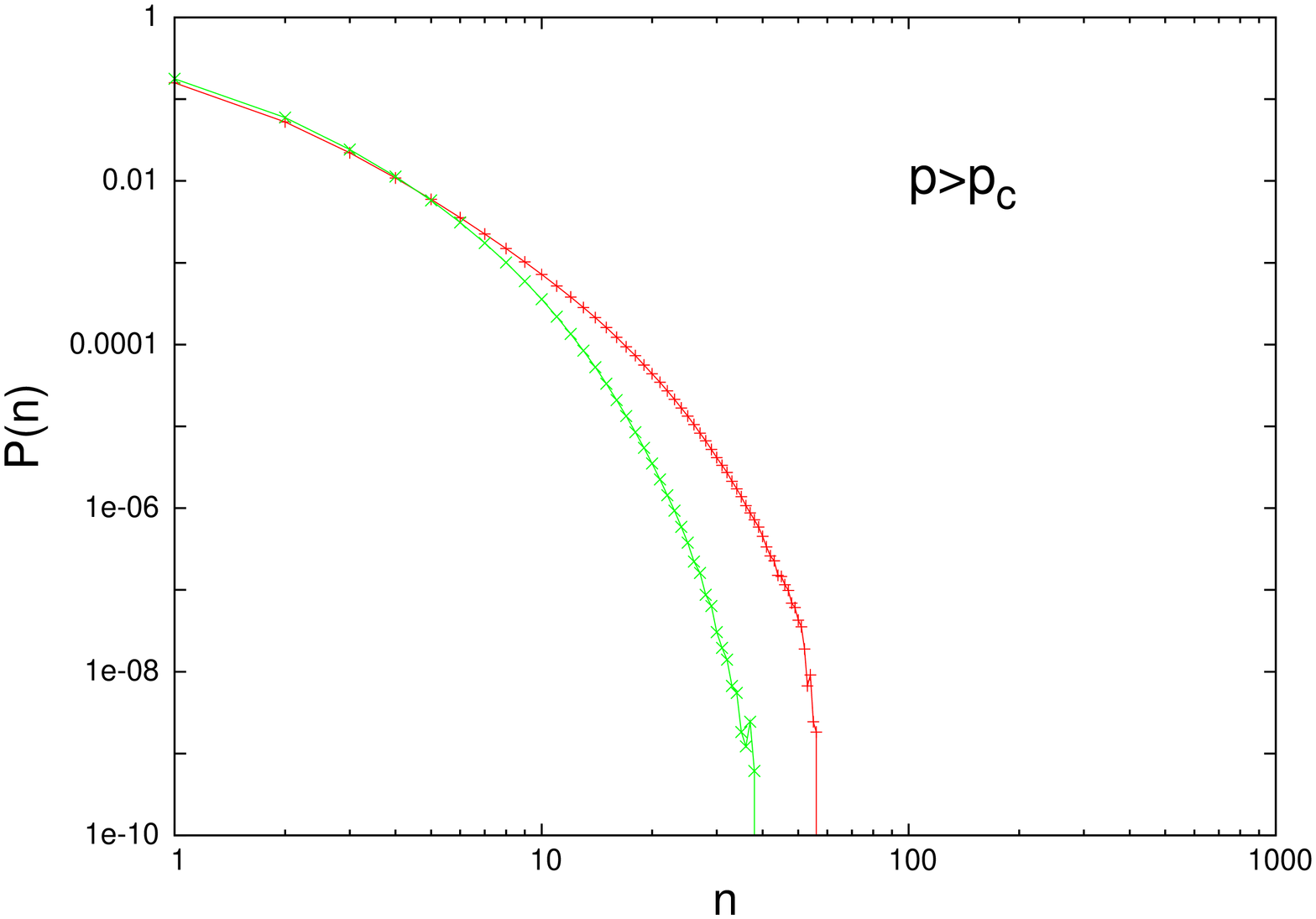}
\caption{Probability distribution $P(n)$ versus $n$ for (a) (top left) $p<p_c$, with $p=0.0$ (red), $0.1$ (green), $0.2$ (blue), (b) (top right) $p=p_c$ where $p_c=0.24$ and the straight line shows a power law with exponent $-2.3$ and (c) (bottom) $p>p_c$ with $p=0.5$ (red), $1.0$ (green).}
\label{probdist}
\end{center}
\end{figure}

We performed Monte-Carlo simulations using the dynamics described in the previous section. We computed the distribution $P(n)$ which gives the probability of finding $n$ particles on any given lattice site. Fig.~(\ref{probdist}) shows the variation of $P(n)$ as a function of the parameter $p$. The number of particles here is $N=\frac{L}{2}-1$ ($L=2048$, $N=1023)$ . For smaller values of $p$ we see a peak in the distribution function at large $n$, in addition to a continuous function at small $n$. The peak represents a condensate while the continuous part represents the distribution of particles which are not part of the condensate. The peak disappears as $p$ is increased beyond a critical value $p_c$ indicating that there is no condensate for $p \ge p_c$. The value of $p_c$ in this case is $0.24$. The probability distribution at $p_c$ is a power law and the exponent $-2.3 \pm 0.1$ seems to best fit the data at small arguments. The data for $p>p_c$ does not fit a simple functional form. The noncondensate part of the probability distribution for $p<p_c$ also does not fit a simple function unlike previously studied models where it is shown to be a power law \cite{satya,satya1}.\\ 

\begin{figure}
\begin{center}
\includegraphics[width=0.51 \linewidth,angle=-90]{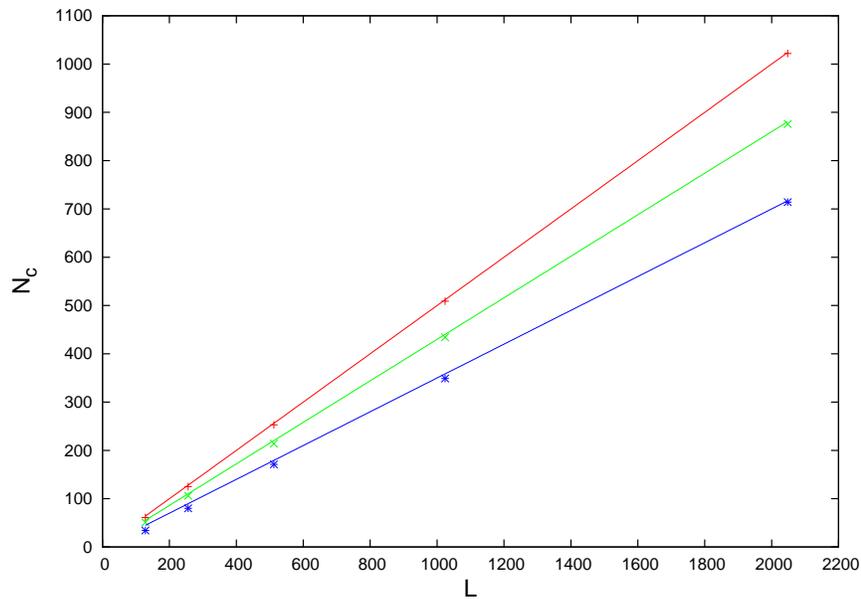}
\caption{Average number of particles in the condensate, $N_c$, plotted versus the system size $L$ for $p=0$(red), $0.1$(green), $0.16$ (blue). The total number of particles is $N=\frac{L}{2}-1$. The slope of the straight lines indicates that the fraction of mass contained in the condensate, $\frac{N_c}{N}$, is $1$, $0.86$ and $0.7$ for $p=0$, $0.1$ and $0.16$ respectively.}
\label{cluster}
\end{center}
\end{figure}

Figure~(\ref{cluster}) plots the average number of particles in the condensate, $N_c$, as a function of the system size $L$, for density close to half ($N=\frac{L}{2}-1$). The data fits a straight line indicating that the number of particles in the condensate is a finite fraction of the total number. This fraction can be inferred from the slope of the lines. As expected from the ZRP results, at $p=0$ nearly all of the mass is contained in the condensate. As $p$ is increased, the fraction of particles in the condensate decreases but is still a finite fraction of the total mass. Thus the picture of the steady state in the $0<p<p_c$ regime is that in addition to a condensate, there is a finite number of particles distributed homogeneously on the lattice.\\ 

\begin{figure}
\begin{center}
\includegraphics[width=0.51 \linewidth,angle=-90]{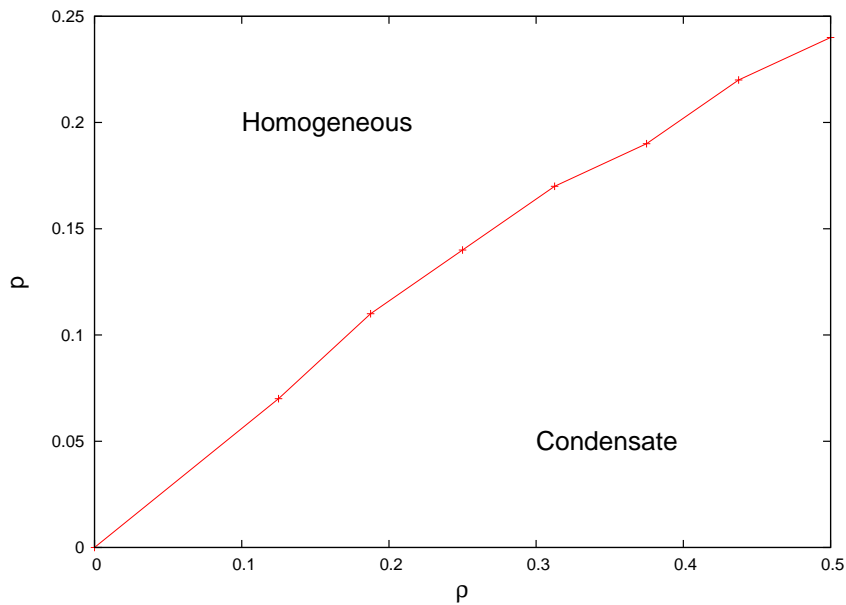}
\caption{The figure shows the phase boundary between the condensate and the homogeneous density phases. The x-axis shows the density $\rho=\frac{N}{L}$ and the $y-axis$ shows $p$.}
\label{phase}
\end{center}
\end{figure}

We computed the probability distribution $P(n)$ for different values of particle density $\rho~(=N/L)$ and studied the change in steady state as a function $p$. Our results show that there is a phase transition and the value of $p_c$ decreases with the decrease in density. The phase diagram in Fig.~(\ref{phase}) sums up our results. While we have kept the density constant and studied the phase change as a function of $p$, one could equivalently see the transition from a homogeneous density to a condensate phase by keeping $p$ constant and increasing the density.  

\section{Discussion}

The factorisation of steady state probability in the zero range process makes analytic treatment possible. The criterion for factorisability has been discussed in~\cite{evans} and it has been shown that the steady state factorises if the hop rate from a site $l$ to a site $k$, $u_{kl}(n_l )$, has the general form $u_{kl}(n_l)=u_l(n_l )W_{kl}$. Here $u_l(n_l)$ gives the probability of a particle leaving the site $l$ which is occupied by $n_l$ particles and $W_{kl}$ is the probability that the particle hops to site $k$. In our case however, there is an explicit dependence of $W_{kl}$ on $n_l$ and the steady state in our model can not be written in a factorised form. When exact treatment is not possible, mean-field analysis has proven to be successful in studying phase transitions in similar systems~\cite{satya,levine}. We will show that in our case, a mean-field equation that assumes no correlation between the sites does not capture the phase transition and the steady state behaviour. We will provide a heuristic understanding of the phase transition mechanism and discuss a possible cause of the failure of the mean-field treatment.\\

 We can write the equation for the probability $P(n)$ in the mean-field approximation as

\begin{eqnarray}
\frac{dP(n)}{dt} &=& -P(n)u(n)-P(n)\left( (1-p)\sum_{k=1}^{N-n}P(k)u(k)+p\sum_{k=1}^{N-n}P(k)u(k) \right) \\
& & +P(n+1)u(n+1)+P(n-1)\left((1-p)\sum_{k=1}^{N-n}P(k)u(k)+p\sum_{k=1}^{N-n}P(k)u(k)\right) \nonumber
\label{meanfield1}
\end{eqnarray}

Here $u(n)=1/n$ is the probability of a particle escaping a site having $n$ particles. The probability $P(n)$ will decrease if a site occupied with $n$ particles either loses or gains a particle. The first term on the RHS is a loss term arising from a particle leaving a site having $n$ particles. The second term accounts for a particle being added to the $n$ particles already present. This may happen in two ways; via a local hop from the nearest neighbours or from a non-local hop from any site on the lattice. Thus the term with co-efficient $(1-p)$ inside the bracket comes from the local jumps and the summation is over all possible occupancies for the nearest neighbour site. The term inside the bracket with co-efficient $p$ arises from non-local hops, the summation here is over sites at a distance $k$ which have exactly $k$ particles. The third and fourth terms are gain terms - $P(n)$ increases if one of the $n+1$ particles present at a site leaves or if a particle enters a site having $n-1$ particles. The last term again has local and non-local contributions. In all terms, there is an implied left and right contribution of $1/2$ and $1/2$ each, which add up to give $1$.\\

The above equation reduces to 

\begin{eqnarray}
\frac{dP(n)}{dt} &=& -P(n)u(n)-P(n)\sum_{k=1}^{N-n}P(k)u(k)+P(n+1)u(n+1) \nonumber \\
 & & +P(n-1)\sum_{k=1}^{N-n}P(k)u(k)
\label{meanfield2}
\end{eqnarray}
which is independent of $p$ and carries no information about the non-local dynamics. In fact Eq.~(\ref{meanfield2}) is the same as that for a simple ZRP with nearest neighbour hops i.e. the p=0 limit of our system. It therefore describes a steady state with a condensate containing all the particles. Thus, the above equation does not describe phase transition in the one dimensional system we are looking at. Since mean field theory is expected to work in higher dimensions, the observed transition should be a feature restricted to low dimensions.\\

A comparison of the relevant time scales in a system can provide insight into the cause of the phase transition~\cite{rajesh} and we employ a similar method to analyse our system. Consider the system to have a condensate of mass $n'$ at some value of $p=p'$. Since the rate of a particle to leave the condensate is $u(n')=1/n'$, one particle is lost every $n'$ time steps. A fraction $p$ of these particles make non-local hops, therefore there is one non-local hop in $n'/p'$ time steps. Now consider the time of return to the condensate of the particle that has made a non-local hop. Since the density on the lattice is small, this particle will return diffusively, making mainly local hops. Thus the time of return of this particle goes as $n'^2$. This is comparable to the time it would take for the whole condensate to melt down, $t'=\frac{1}{p'}\sum_{i=1}^{n'} i \sim n'^2$. The local hops are irrelevant as the particles can return to the condensate in a time of order one and therefore cannot cause declustering.\\

So we see that it is an interplay of the two time scales of condensate melting versus diffusive return that determines the phase of the system. Which one of these dominates is determined by the value of $p$ as well as the density of particles on the lattice. A higher value of $p$ leads to more declustering and a smaller density leads to smaller condensates, both of which favour a homogeneous density phase. On the other hand a higher density and a smaller value of $p$ will favour condensate formation. This heuristic argument, if true, leads to interesting consequences which serve as checks for its validity.\\

\begin{figure}
\begin{center}
\includegraphics[width=0.6 \linewidth,angle=-90]{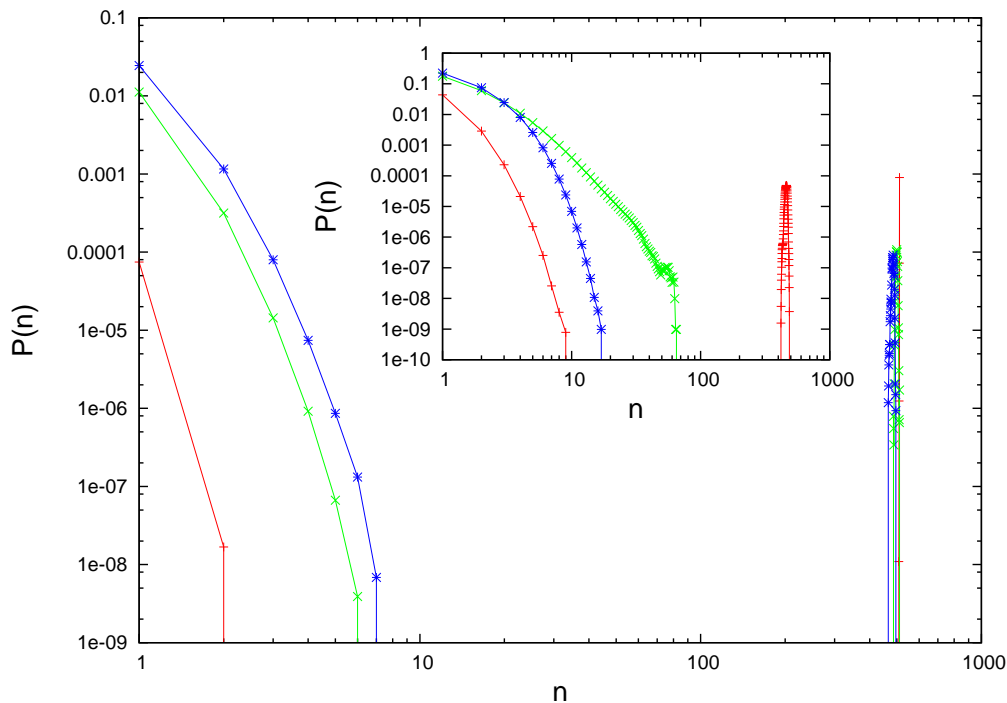}
\caption{The main figure shows $P(n)$ versus $n$ for $\alpha=1.5$ and $p=0.0$ (red), $0.5$ (green), $1.0$ (blue). The condensate persists for all values of $p$. The inset shows $P(n)$ versus $n$ for $\alpha=0.5$ and $p=0.0$ (red), $0.03$ (green), $1.0$ (blue). There is a condensate phase for $p=0$ but it vanishes at a very small value of $p$.}
\label{alphadep}
\end{center}
\end{figure}

Suppose the probability of a particle leaving a site has a general form $u(n)=1/n^{\alpha}$, with our model above corresponding to the special case of $\alpha=1$. The argument of competing time scales predicts that for $\alpha>1$, the steady state will have a condensate for any value of $p$. This is borne out in our simulations, Fig.~(\ref{alphadep}). For $\alpha<1$, the $p=0$ state is a condensate phase as predicted by the ZRP but we expect that there should be no condensate for $p>0$. We indeed see that for small $\alpha$, the condensate disappears at very small values of $p$, Fig.~(\ref{alphadep}) (inset). Another possible alteration in our original model can be that the non-local hop, instead of having a length $n$, has a length equal to the integer part of $n^{\gamma}$. For $\gamma < 1$, we expect from the argument of time scales that the condensate will always persist. We have also confirmed this prediction with our numerics. These results give credence to our argument.\\

\begin{figure}
\begin{center}
\includegraphics[width=0.6 \linewidth,angle=-90]{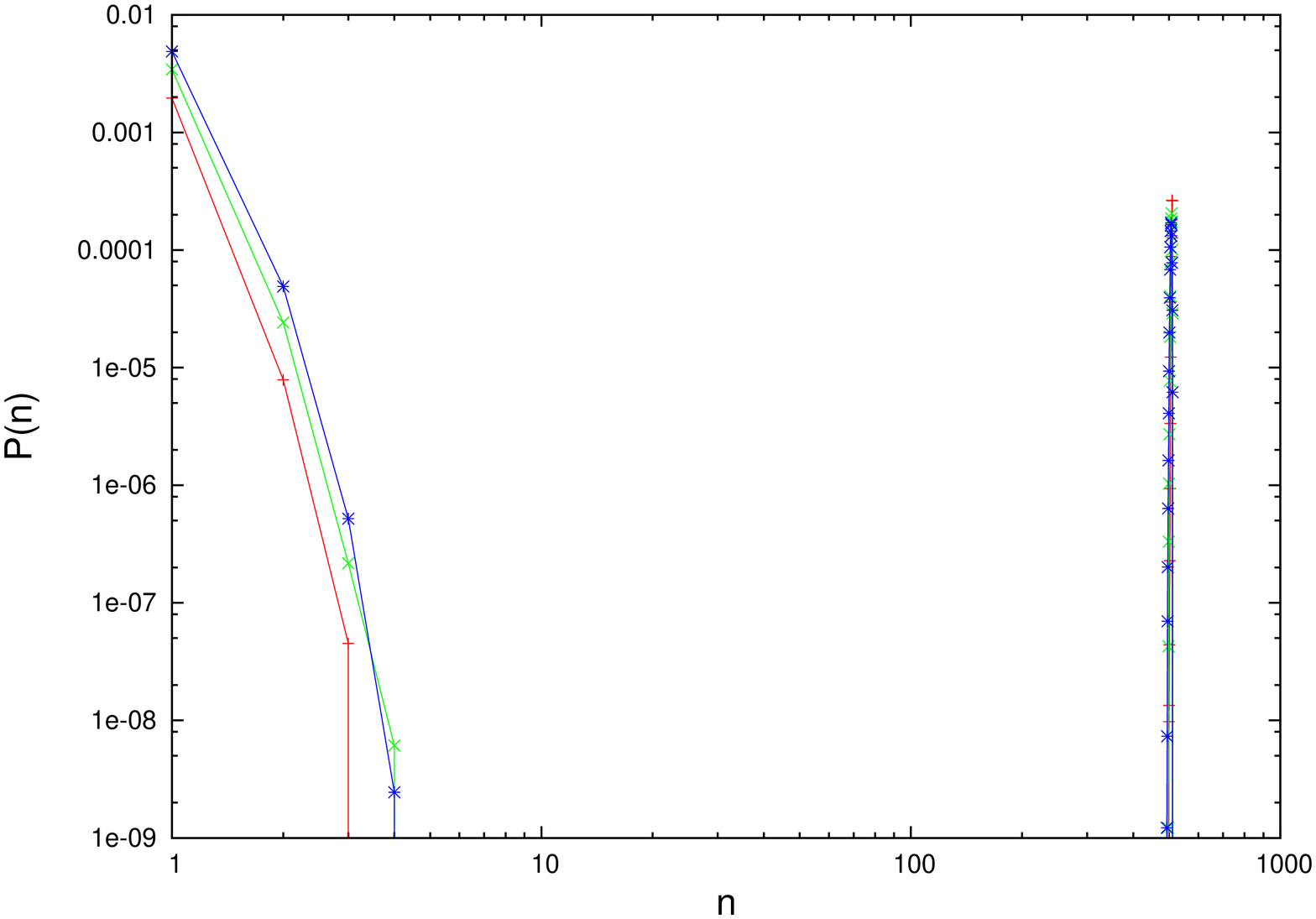}
\caption{$P(n)$ versus $n$ for a model with asymmetric dynamics. The particles preferably move to the right (probability $0.6$) rather than symmetrically (probability $0.5$ on each side.). We show data for $p=0.0$ (red), $0.5$ (green) and $1.0$ (blue). We see that the condensate exists for all values of $p$. The system size $L=1024$ and the number of particles $N=511$.}
\label{asymmetry}
\end{center}
\end{figure}

We also studied the effect of introducing asymmetry in the dynamics. In the model above, the particles have an equal probability of moving either in the left or the right direction. We introduced an asymmetry so that the probability of moving in one direction is greater than the other. Because of this asymmetry, particles making a non-local hop can return to the cluster in a time proportional to size of the cluster $n'$ instead of the diffusive $n'^2$. Thus the asymmetry acts to stabilise the condensate. Our results in Fig.~(\ref{asymmetry}) show that the introduction of even a slight asymmetry causes the homogeneous density phase to disappear. We again see that it is the diffusive dynamics which is responsible for the phase transition.\\ 

Let us now consider dimensions higher than one. Since we are working with a finite density, the number of particles on the lattice scales as $L^d$, $d$ being the dimension. Thus the size of the condensate also scales as $L^d$. On the other hand, our dynamical rules do not allow the particles to make jumps of length larger than the system size $L$. Thus a particle farthest from the condensate will make a diffusive return in time of order $L^2$. This is always smaller than the time scale of cluster melting which goes faster than $L^{d+1}$. Thus, if we define a restricted model in higher dimension, it will always have a condensate as predicted by the mean field theory. Our model therefore is expected to work only in one dimension.

\section{Conclusions}

We have studied phase transition in a nonequilibrium lattice model in one dimension. The transition from a condensate phase to a homogeneous density phase is caused by two opposing factors - the attractive interaction between the particles which promotes condensation and the non-local move which destabilises the condensate. We find that the diffusive dynamics is crucial to the transition, as is the fact that the leaving probability $u(n)$ is inversely proportional to the number of particles in the cluster, $u(n)=1/n$. This explains why the transition takes place only in one dimension and the mean-field treatment fails. For the same reason, an introduction of asymmetry in the dynamics leads to only one phase, the condensate phase. 

\section{Acknowledgements}

The author would like to thank Kavita Jain for useful comments.

\end{document}